# A Modified Design Of Acf Operation For Reducing Papr Of Ofdm Signal

Md. Munjure Mowla[1], Md. Yeakub Ali[2] and S.M. Mahmud Hasan[3]

[1,2,3] Department of Electronics & Telecommunication Engineering
Rajshahi University of Engineering & Technology, Rajshahi, Bangladesh

*ABSTRACT*

*Next generation wireless communication technology long term evolution (LTE) has implemented orthogonal frequency division multiplexing (OFDM) technique as a strong candidate for radio access systems. It has several attributes such as providing robustness to multipath fading & impulse noise, eliminating intersymbol interference (ISI), inter carrier interference (ICI) & the need for equalizers. The major challenging issue of OFDM technique is the high peak to average power ratio (PAPR) which is defined as the ratio of the peak power to the average power of the OFDM signal. A trade-off is necessary for reducing PAPR with increasing bit error rate (BER), computational complexity or data rate loss etc. In this paper, a moderately modified design of amplitude clipping & filtering operation (ACF) is proposed and implemented which shows the significant improvement in case of PAPR reduction for both quadrature phase shift keying (QPSK) & quadrature amplitude modulation (QAM) while increasing slight BER match up to to an existing method.*

*KEYWORDS*

*Bit Error rate (BER), Complementary Cumulative Distribution Function (CCDF), Long Term Evolution (LTE), Orthogonal Frequency Division Multiplexing (OFDM) and Peak to Average Power Ratio (PAPR).*

## 1. INTRODUCTION

Third generation partnership project (3GPP) adopted next generation wireless communication technology Long Term Evolution (LTE) is designed to increase the capacity and speed of existing mobile telephone & data networks. LTE has adopted a multicarrier transmission technique known as orthogonal frequency division multiplexing (OFDM). OFDM meets the LTE requirement for spectrum flexibility and enables cost-efficient solutions for very wide carriers [1]. The additional increasing demand on high data rates in wireless communications systems has arisen in order to carry broadband services. OFDM offers high spectral efficiency, immune to the multipath fading, low inter-symbol interference (ISI), immunity to frequency selective fading and high power efficiency. Today, OFDM is used in many emerging fields like wired Asymmetric Digital Subscriber Line (ADSL), wireless Digital Audio Broadcast (DAB), wireless Digital Video Broadcast - Terrestrial (DVB - T), IEEE 802.11 Wireless Local Area Network (WLAN), IEEE 802.16 Broadband Wireless Access (BWA), Wireless Metropolitan Area Networks (IEEE 802.16d), European ETSI Hiperlan/2 etc [2].

One of the key problems of OFDM is high peak to average power ratio (PAPR) of the transmit signal. If the peak transmit power is limited by application constraints, the effect is to reduce the average power allowed under multicarrier transmission compare to that under constant power modulation techniques. This lessens the range of multicarrier transmission. Furthermore, the transmit power amplifier must be operated in its linear region (i.e., with a large input back-off), where the power conversion is inefficient to avoid spectral growth of the multicarrier signal in the form of intermodulation among subcarriers and out-of-band radiation. It may cause a detrimental effect on battery lifetime in some mobile applications. As handy devices have a finite battery life, it is considerable to find ways of reducing the PAPR allowing for a smaller more efficient high power amplifier (HPA), which in turn will mean a more lasting battery life. In many low-cost applications, the problem of high PAPR may outweigh all the potential benefits of multicarrier transmission systems [3].

A number of promising approaches or processes have been proposed & implemented to reduce PAPR with the expense of increase transmit signal power, bit error rate (BER) & computational complexity and loss of data rate, etc.  So, a system trade-off is required. These techniques include Amplitude Clipping and Filtering, Peak Windowing, Peak Cancellation, Peak Reduction Carrier, Envelope Scaling, Decision-Aided Reconstruction (DAR), Coding, Partial Transmit Sequence (PTS), Selective Mapping (SLM), Interleaving, Tone Reservation (TR), Tone Injection (TI), Active Constellation Extension (ACE), Clustered OFDM, Pilot Symbol Assisted Modulation, Nonlinear Companding Transforms etc [4].

## 2. CONCEPTUAL MODEL OF OFDM SYSTEM

OFDM is a special form of multicarrier modulation (MCM) with densely spaced subcarriers with overlapping spectra, thus allowing multiple-access.MCM works on the principle of transmitting data by dividing the stream into several bit streams, each of which has a much lower bit rate and by using these sub-streams to modulate several carriers[5].

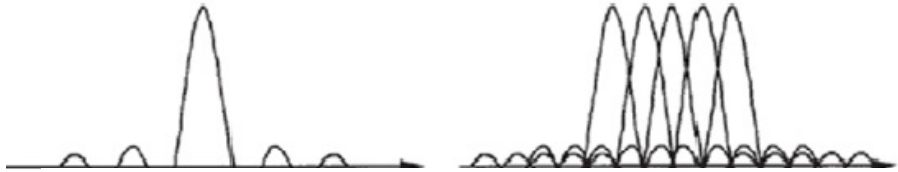

Figure 1.  Spectra of (a) An OFDM Sub-channel and (b) An OFDM Signal [2]

In multicarrier transmission, bandwidth divided in many non-overlapping subcarriers but not essential that all subcarriers are orthogonal to each other as shown in figure 1 (a) [2]. In OFDM the sub-channels overlap each other to a certain extent as can be seen in figure 1 (b), which directs to a resourceful use of the total bandwidth. The information sequence is mapped into symbols, which are distributed and sent over the N sub-channels, one symbol per channel. To allow dense packing and still ensure that a minimum of interference between the sub-channels is met, the carrier frequencies must be chosen carefully.  By using orthogonal carriers, frequency domain can be viewed so as the frequency space between two sub-carriers is given by the distance to the first spectral null [2].

### 2.1. Mathematical Explanation of OFDM Signals

In OFDM systems, a defined number of consecutive input data samples are modulated first (e.g, QPSK or QAM), and then jointly correlated together using inverse fast Fourier transform (IFFT) at the transmitter side [5]. IFFT is used to produce orthogonal data subcarriers. Let, data block of length $N$ is represented by a vector, $X=[X_0, X_1......X_{N-1}]^T$. Duration of any symbol $X_K$ in the set $X$ is $T$ and represents one of the sub-carriers  set. As the N sub-carriers chosen to transmit the signal are orthogonal, so we can have, $f_n = n\Delta f$, where $n\Delta f = 1/NT$ and $NT$ is the duration of the OFDM data block $X$. The complex data block for the OFDM signal to be transmitted is given by [3],

$$x(t) = \frac{1}{\sqrt{N}} \sum_{n=0}^{N-1} X_n e^{j2\pi n \Delta f t} \qquad 0 \leq t \leq NT \qquad (1)$$

Where,

$j = \sqrt{-1}$, $f$ is the subcarrier spacing and *NT* denotes the useful data block period. In the OFDM technique, the subcarriers are definitely chosen as orthogonal characteristics (i.e., $f = 1/NT$). On the other hand, OFDM output symbols typically have large dynamic envelope range due to the superposition procedure performed at the IFFT stage in the transmitter stage.

## 3. BASICS OF PAPR

Due to the incidence of large number of independently modulated sub-carriers in an OFDM system, the peak value of the system may be very high as compared to the average of the total system. The coherent summation of N signals of same phase produces a peak which is N times the average signal [3]. PAPR is an vital factor in the design of both high power amplifier (HPA) and digital-to-analog (DAC) converter, for generating almost error-free (minimum errors) transmitted OFDM symbols. So, the ratio of peak power to average power is known as PAPR.

$$PAPR = \frac{Peak\_Power}{Average\_Power}$$

The PAPR of the transmitted signal is defined as [6],

$$PAPR[x(t)] = \frac{\max_{0 \leq t \leq NT} |x(t)|^2}{P_{av}} = \frac{\max_{0 \leq t \leq NT} |x(t)|^2}{\frac{1}{NT} \int_0^{NT} |x(t)|^2 \, dt} \qquad (2)$$

Where, $P_{av}$ is the average power of and it can be computed in the frequency domain because Inverse Fast Fourier Transform (IFFT) is a (scaled) unitary transformation.

For superior estimated the PAPR of continuous time OFDM signals, the OFDM signals samples are obtained by *L* times oversampling [3]. *L* times oversampled time domain samples are *LN* point IFFT of the data block with *(L-1)N* zero-padding. Therefore, the oversampled IFFT output can be expressed as,

$$x[n] = \frac{1}{\sqrt{N}} \sum_{k=0}^{N-1} X_k e^{j2\pi nk/LN} \qquad 0 \leq n \leq NL - 1 \qquad (3)$$

### 3.1. Objectives behind PAPR reduction

The difficulties associated with OFDM are inherited by OFDMA (Orthogonal frequency division multiple access) technique. OFDMA also suffers from high peak to average power ratio (PAPR) because it is inherently made up of so many subcarriers [7]. The subcarriers are summed up constructively to form large peaks. Hence, high peak power requires High Power Amplifiers (HPA), ADC and DAC converters. Most of the wireless systems employ the HPA in the transmitter to obtain enough transmission power. For the proposed of achieving the maximum output power efficiency, the HPA is usually operated at or near the saturation region [6].

The efficiency of power is pivotal in wireless communication as it provides sufficient area coverage, saves power consumption and allows small size terminals etc. It is important to target at a power efficient operation of the non-linear HPA with low back-off values and try to give possible solutions to the interference problem. In addition, the non- linear characteristic of the HPA is very approachable to the variation in signal amplitudes. The variation of OFDM signal

33

amplitudes is very wide with high PAPR. So, HPA will introduce inter-modulation between the different subcarriers and create extra interference into the systems due to high PAPR of OFDM signals. This additional interference forwards to BER increment. To reduce the signal distortion and keep a minimum BER, it requires a linear work in its linear amplifier region with a large dynamic range. Actually, this linear amplifier has poor efficiency and is so expensive. Therefore, a better solution is to prevent the occurrence of such interference by reducing the PAPR of the transmitted signal with some manipulations of the OFDM signal itself [6].

Large PAPR also demands the DAC with large dynamic range to accommodate the large peaks of the OFDM signals. Although, a high precision DAC supports high PAPR with a reasonable amount of quantization noise, but it might be very expensive for a given sampling rate of the system. Moreover, OFDM signals show Gaussian distribution for large number of subcarriers, which means the peak signal quite rarely occur and uniform quantization by the ADC is not desirable. If the peak signal is clipped, it will introduce in-band distortion (BER increment) and out-of-band radiation (adjacent channel interference) into the channel. It is therefore the perfect solution to reduce the PAPR before OFDM signals are transmitted into nonlinear HPA and DAC.

## 4. CONVENTIONAL CLIPPING AND FILTERING

Amplitude Clipping and Filtering is one of the easiest techniques which may be under taken for PAPR reduction for an OFDM system. A threshold value of the amplitude is fixed in this case to limit the peak envelope of the input signal [8].

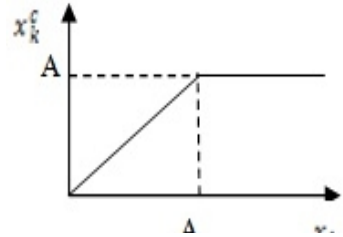

Figure 2. Clipping Function

The clipping ratio (CR) is defined as,

$$CR = \frac{A}{\sigma} \qquad (4)$$

Where, A is the amplitude and $\sigma$ is the root mean squared value of the unclipped OFDM signal. The clipping function is performed in digital time domain, before the D/A conversion and the process is described by the following expression,

$$x_k^c = \begin{cases} x_k & |x_k| \le A \\ Ae^{j\phi(x_k)} & |x_k| > A \end{cases} \qquad 0 \le k \le N-1 \qquad (5)$$

Where, $x_k^c$ is the clipped signal, $x_k$ is the transmitted signal, A is the amplitude and $\phi(x_k)$ is the phase of the transmitted signal $x_k$.

34

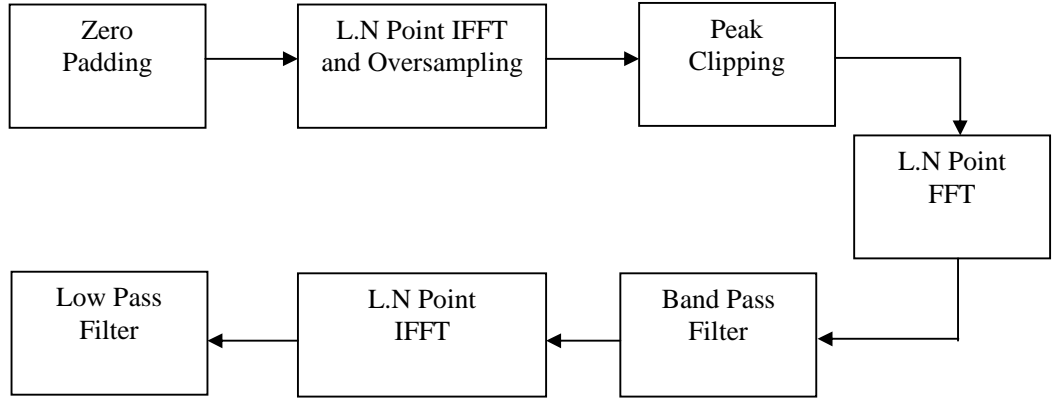

Figure 3. Block Diagram of PAPR reduction using Conventional Clipping & Filtering [9]

In the figure 3, the conventional block diagram of clipping & filtering method is shown which have some limitations described in below.

### 4.1. Limitations of Clipping and Filtering

➢ Clipping causes in-band signal distortion, therefore BER performance is degraded.

➢ Another problem of clipping is out-of-band radiation (channel interference), which imposes out-of-band interference signals to adjacent channels. Though the out-of-band radiation caused by clipping can be reduced by filtering, it may affect high-frequency components of in-band signal (aliasing) when the clipping is performed with the Nyquist sampling rate in the discrete-time domain [9].

➢ On the other hand, if clipping is performed for the sufficiently-oversampled OFDM signals (e.g., L  4) in the discrete-time domain before a low-pass filter (LPF) and the signal passes through a band-pass filter (BPF), the BER performance will be less degraded [9].

➢ Filtering the clipped signal can reduce out-of-band radiation at the cost of peak regrowth. The signal after filtering operation may exceed the clipping level specified for the clipping operation [3].

## 5. PROPOSED CLIPPING AND FILTERING SCHEME

As our main focus is to reduce PAPR, so, in this simulation, we have trade-off between PAPR reduction with BER increment. Very less amount of BER increment is desirable. Pointing out the third limitation in section 4.1, (Yong Soo et.al) [9] shows that if clipped signal passes through a Composed filter (FIR based BPF) before passing a LPF to reduce out-of-band radiation, then it causes less BER degradation with medium amount of PAPR reduction.

Considering this concept, we design a scheme for clipping & filtering method where clipped signal passes through a Composed filter (IIR based BPF) before passing a LPF, then it causes a little bit more BER degradation but more amount of PAPR reduction than existing method[9].
This proposed scheme is shown in the figure 4. It shows a block diagram of a PAPR reduction scheme using clipping and filtering, where *L* is the oversampling factor and *N* is the number of subcarriers. The input of the IFFT block is the interpolated signal introducing *N(L −1)* zeros (also, known as zero padding) in the middle of the original signal is expressed as,

35

$$X'[k] = \begin{cases} X[k], & \text{for } 0 \leq k \leq \frac{N}{2} \text{ and } NL - \frac{N}{2} < k < NL \\ 0 & \text{elsewhere} \end{cases} \quad (6)$$

In this system, the L-times oversampled discrete-time signal is generated as,

$$x'[m] = \frac{1}{\sqrt{LN}} \sum_{k=0}^{LN-1} X'[k] e^{\frac{j2\pi m k}{LN}}, \qquad m = 0,1,...NL-1 \quad (7)$$

and is then modulated with carrier frequency $f_c$ to yield a passband signal $x^p[m]$.

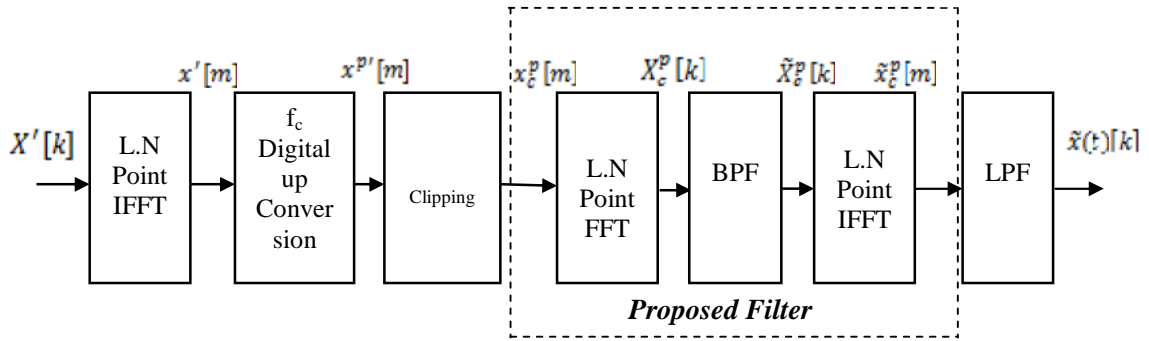

Figure 4. Block Diagram of Proposed Clipping & Filtering Scheme.

Let $x_c^p[m]$ denote the clipped version of $x^p[m]$, which is expressed as,

$$x_c^p[m] = \begin{cases} -A & x^p[m] \leq -A \\ x^p[m] & |x^p[m]| < A \\ A & x^p[m] \geq A \end{cases} \quad (8)$$

Or,

$$x_c^p[m] = \begin{cases} x^p[m] & \text{if, } |x^p[m]| < A \\ \frac{x^p[m]}{|x^p[m]|} \cdot A & \text{otherwise} \end{cases} \quad (9)$$

Where, *A* is the pre-specified clipping level. After clipping, the signals are passed through the Composed filter *(Proposed Filter).* This composed filter itself consists on a set of FFT-IFFT operations where filtering takes place in frequency domain after the FFT function. The FFT function transforms the clipped signal $x_c^p[m]$ to frequency domain yielding $X_c^p[k]$. The information components of $X_c^p[k]$ are passed to a band pass filter (BPF) producing $\tilde{X}_c^p[k]$. This filtered signal is passed to the unchanged condition of IFFT block and the out-of-band radiation that fell in the zeros is set back to zero. The IFFT block of the filter transforms the signal to time domain and thus obtain $\tilde{x}_c^p[m]$.

## 6. DESIGN AND SIMULATION PARAMETERS

In this simulation, an IIR digital filter (Chebyshev Type I) is used in the composed filtering. Chebyshev Type I filter is equiripple in the passband and monotonic in the stopband. Type I filter rolls off faster than type II filters. Chebyshev filter has the property that it minimizes the error between the idealized and the actual filter characteristic over the range of the filter. Because of the passband equiripple behaviour inherent in Chebyshev Type I filter, it has a smoother response. Using the special type of bandpass filter in the composed filter, significant improvement is observed in the case of PAPR reduction. Table 1 shows the values of parameters used in the QPSK & QAM system for analyzing the performance of clipping and filtering technique [9]. We have simulated the both methods (Existing and Proposed) with the same parameters at first. But, simulation results show the significant improvement occurs in PAPR reduction for proposed method.

Table 1. Parameters Used for Simulation of Clipping and Filtering.

| Parameters | Value |
| --- | --- |
| Bandwidth ( BW) | 1 MHz |
| Over sampling factor (L) | 8 |
| Sampling frequency, $f_s$ = BW*L | 8 MHz |
| Carrier frequency, $f_c$ | 2 MHz |
| FFT Size / No. of Subscribers (N) | 128 |
| CP / GI size | 32 |
| Modulation Format | QPSK and QAM |
| Clipping Ratio (CR) | 0.8, 1.0, 1.2, 1.4, 1.6 |

### 6.1. Simulation Results for PAPR Reduction

At first, we simulate the PAPR distribution for CR values =0.8, 1.0, 1.2, 1.4, 1.6 with QPSK modulation and N=128. Then, we simulate with QAM modulation and N=128 and compare different situations.

### 6.1.1 Simulation Results: (QPSK and N=128)

In the existing method, PAPR distribution for different CR value is shown in figure 5 (a). At CCDF =$10^{-1}$, the unclipped signal value is 13.46 dB and others values for different CR are tabulated in the table 2.

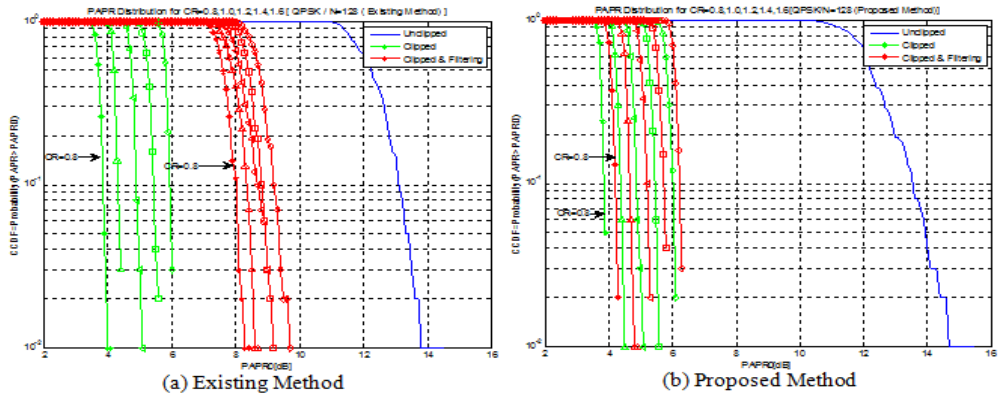

Figure 5. PAPR Distribution for CR=0.8,1.0,1.2,1.4,1.6 [QPSK and N=128]

In the proposed method, simulation shows the significant reduction of PAPR is shown for different CR values in figure 5(b). At CCDF =$10^{-1}$, the unclipped signal value is 13.52 dB and others values for different CR are tabulated in the table 2. From table 2, it is clearly observed that the proposed method reduces PAPR significantly with respect to existing Yong Soo [9] analysis. Proposed method of filter design is done with the same parameters that used in [9].

Table 2. Comparison of Existing with Proposed Method for PAPR value [QPSK and N=128]

| CR value | PAPR value (dB) (Existing) | PAPR value (dB) (Proposed) | Improvement in PAPR value (dB) |
|---|---|---|---|
| 0.8 | 8.10 | 4.21 | 3.89 |
| 1.0 | 8.35 | 4.67 | 3.68 |
| 1.2 | 8.72 | 5.21 | 3.51 |
| 1.4 | 8.81 | 5.72 | 3.09 |
| 1.6 | 9.22 | 6.29 | 2.93 |

### 6.1.2 Simulation Results: (QAM and N=128)

The simulation results are now shown for QAM modulation and no. of subscribers, N=128. In the existing method, PAPR distribution for different CR value is shown in figure 6 (a). At CCDF =$10^{-1}$, the unclipped signal value is 13.42 dB and others values for different CR are tabulated in the table 3. In the Proposed method, simulation shows the significant reduction of PAPR is shown for different CR values in figure 6(b). At CCDF =$10^{-1}$, the unclipped signal value is 13.65 dB and others values for different CR are tabulated in the Table 3.

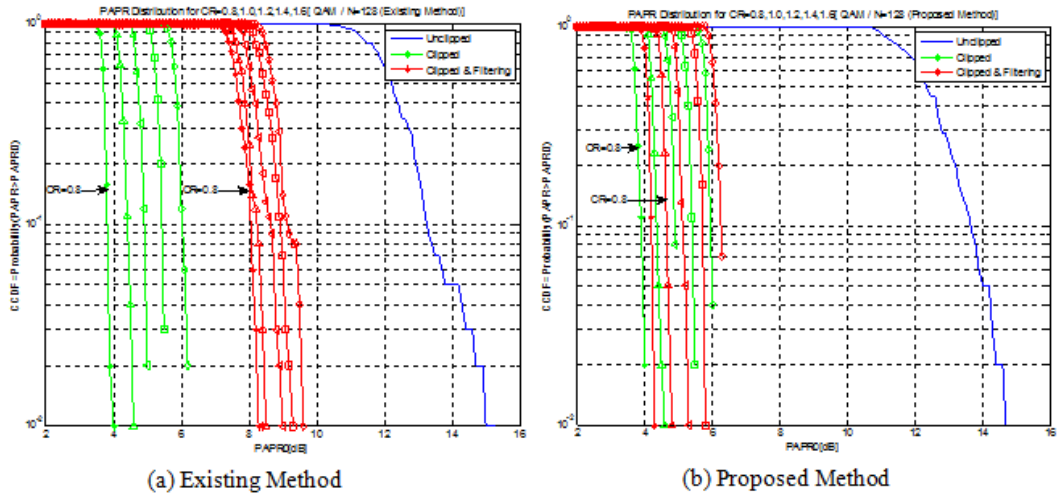

(a) Existing Method  (b) Proposed Method

Figure 6. PAPR Distribution for CR=0.8,1.0,1.2,1.4,1.6 [QAM and N=128]

The existing method and proposed method PAPR distribution values for different CR values are tabulated and differences are shown in table 3. From table 3, it is clearly observed that the

International Journal of Next-Generation Networks (IJNGN) Vol.6, No.1, March 2014

proposed method reduces PAPR significantly with respect to existing Yong Soo [9] analysis for QAM and N=128 also. So, proposed method works on efficiently for both QPSK & QAM.

Table 3. Comparison of Existing with Proposed Method for PAPR value [QAM and N=128]

| CR value | PAPR value (dB) (Existing) | PAPR value (dB) (Proposed) | Improvement in PAPR value (dB) |
|---|---|---|---|
| 0.8 | 8.23 | 4.21 | 4.02 |
| 1.0 | 8.32 | 4.65 | 3.67 |
| 1.2 | 8.65 | 5.11 | 3.54 |
| 1.4 | 8.92 | 5.71 | 3.21 |
| 1.6 | 9.15 | 6.27 | 2.88 |

Now, if we compare the values for different CR values in case of QPSK & QAM to show the effect of modulation on proposed filter design, it is observed that for the same number of subscribers (N=128) & low CR=0.8, there is no differences between using QAM & QPSK. But, with the increasing value of CR, QAM provides less PAPR than QPSK. So, for high CR, QAM is more suitable than QPSK in case of proposed filter.

## 6.2. Simulation Results for BER Performance

The clipped & filtered signal is then passed through the AWGN channel and BER are measured for both existing & proposed methods. We have also simulated the analytical BER curve that is shown in the curve. Figure 7 and figure 8 show the BER performance for QPSK and QAM with N=128. It is seen from these figures that the BER increases as the CR decreases.

### 6.2.1 Simulation Results: (QPSK and N=128)

Now, for QPSK & N=128 with all other same data mentioned in table 1, both existing and proposed methods are executed and resulted graphs are shown in figure 7(a) & figure 7(b) respectively. It is observed from these two figures that BER increases slightly in proposed method with respect to existing method for same value of CR.

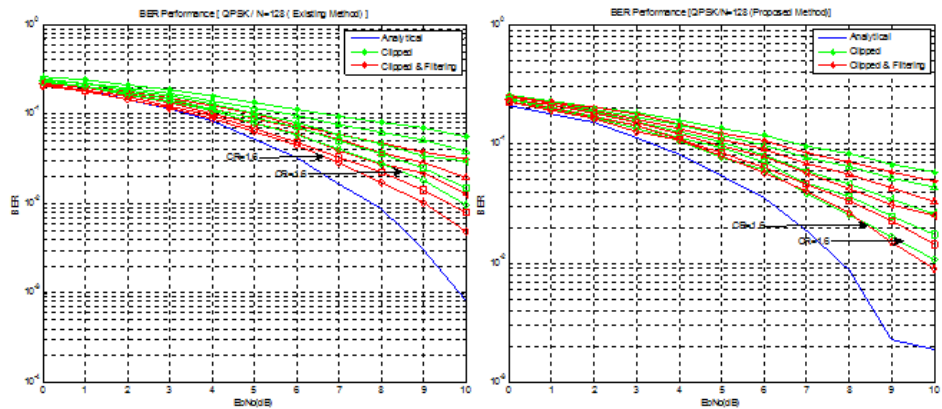

Figure 7. BER Performance [QPSK and N=128]

The measured BER value at 6 dB point is tabulated in table 4.

Table 4. Comparison of BER value for Existing & Proposed Method [QPSK and N=128]

| CR value | BER value (Existing) | BER Value (Proposed) | Difference in BER value |
|---|---|---|---|
| 0.8 | 0.07413 | 0.10631 | -0.03218 |
| 1.0 | 0.06984 | 0.09012 | -0.02028 |
| 1.2 | 0.05982 | 0.07846 | -0.01864 |
| 1.4 | 0.04891 | 0.06358 | -0.01467 |
| 1.6 | 0.04449 | 0.05748 | -0.01299 |

BER performance is measured and compared in both the table 4 (QPSK) & table 5(QAM) which shows different CR values for both existing & proposed method in case of same parameter value.

From table 4, it is observed that, for CR values (0.8,1.0,1.2,1.4 & 1.6) , the difference magnitude between existing & proposed method are 0.3218,0.02028,0.01864,0.01467 & 0.01299 respectively in QPSK. These BER degradations are acceptable as these are very low values.

**6.1.2 Simulation Results: (QAM and N=128)**

Again, for QAM & N=128 with all other same data mentioned in table 1, both existing and proposed methods are executed and resulted graphs are shown in figure 8(a) & figure 8(b) respectively. It is observed from these two figures that BER increases slightly in proposed method with respect to existing method for same value of CR.

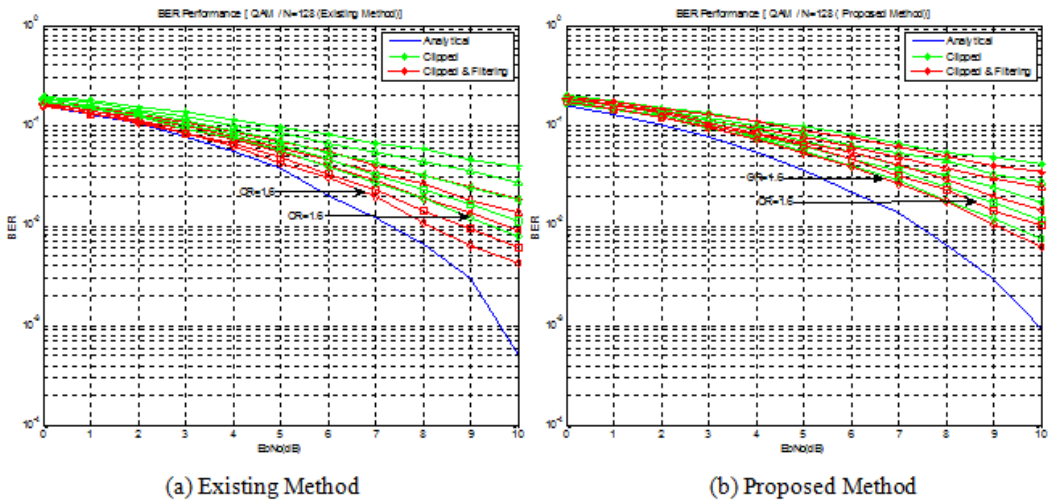

Figure 8. BER Performance [QAM and N=128]

The measured BER value at 6 dB point is tabulated in table 5.

Table 5. Comparison of BER value for Existing & Proposed Method [QAM and N=128]

International Journal of Next-Generation Networks (IJNGN) Vol.6, No.1, March 2014| CR value | BER value (Existing) | BER Value (Proposed) | Difference in BER value |
|---|---|---|---|
| 0.8 | 0.05563 | 0.07535 | -0.01972 |
| 1.0 | 0.04614 | 0.06098 | -0.01484 |
| 1.2 | 0.04016 | 0.05433 | -0.01417 |
| 1.4 | 0.03268 | 0.04631 | -0.01363 |
| 1.6 | 0.03055 | 0.04211 | -0.01156 |

From table 5, it is observed that, for CR values (0.8,1.0,1.2,1.4 & 1.6) , the difference magnitude between existing & proposed method are 0.01972,0.01484,0.01417,0.01363 & 0.01156 respectively in QAM. These BER degradations are acceptable as these are very low values.

Now, if we compare the values for different CR values in case of QPSK & QAM to show the effect of modulation on proposed filter design, it is observed that for the same simulation parameters, QAM provides less BER degradation than QPSK in all cases. With gradual increasing of CR values, the differences of BER values between QPSK & QAM also becomes decreasing. So, QAM is more suitable for proposed method.

## 6. CONCLUSION

In this paper, a comparative scheme of amplitude clipping & filtering based PAPR reduction technique has been analyzed where PAPR reduces significantly compare to an existing method with slightly increase of BER. At first phase, simulation has been executed for existing method with QPSK modulation and number of subscriber (N=128) and then executed for the proposed method for same parameter and observed that PAPR reduces significantly. Next, the simulation has been performed for QAM modulation & N=128 and result shows the considerable improvement in case of QAM also. Then, the proposed method results for both QAM & QPSK modulation with N=128 has been compared. It is observed from that for the same number of subscribers (N=128) & low CR=0.8, there is no differences between using QAM & QPSK. But, with the increasing value of CR, QAM provides less PAPR than QPSK. So, for high CR, QAM is more suitable than QPSK in case of proposed filter. In case of BER, with gradual increasing of CR values, the differences of BER values between QPSK & QAM also becomes decreasing. So, QAM is more suitable for proposed method. In the present simulation study, ideal channel characteristics have been assumed. In order to estimate the OFDM system performance in real world, Multipath Rayleigh fading would be a major consideration in next time. The increase number of subscribers (N) & other modulation parameters could be another assumption.

## REFERENCES

[1]   Eric Dahlman, Stefan Parkvall, and Johan Skold, 4G LTE / LTE-Advanced for Mobile Broadband, United Kingdom : ELSEVIER Academic Press ,2011.
[2]   R. Prasad, OFDM for Wireless Communications Systems. London, Boston: Artech House, Inc, 2004.
[3]   S. H. Han and J. H. Lee, "An overview of peak-to-average power ratio reduction techniques for multicarrier transmission", IEEE Wireless Comm, vol. 12, no.2, pp.56-65, Apr. 2005.
[4]   M.M.Mowla, S.M.A Razzak and M.O.Goni, Improvement of PAPR Reduction for OFDM Signal in LTE System, Germany: LAP Lambert Academic Publishing, 2013.
[5]   M.M.Mowla and S.M.M Hasan, "Performance Improvement of PAPR Reduction for OFDM Signal In LTE System", International Journal of Wireless & Mobile Networks (IJWMN), Vol. 5, No. 4, pp.35-47, doi:10.5121/ijwmn.2013.5403, August 201341

## Authors

**Md. Munjure Mowla** is now working as an Assistant Professor in Electronics & Telecommunication Engineering department of Rajshahi University Engineering & Technology (RUET) since November 2010. He has completed M.Sc Engineering degree in Electrical & Electronic Engineering (EEE) from RUET in May 2013. He has four years telecom job experience in the operators, vendors, ICX etc of Bangladesh telecom market. Mr. Mowla has published several international journals as well as conference papers and three books. He is a member of communication society COMSOC of IEEE, Institutions of Engineers, Bangladesh (IEB) and Bangladesh Electronics Society (BES). His research interest includes advanced wireless communication including LTE, LTE-Advanced, green communication, smart grid communication.

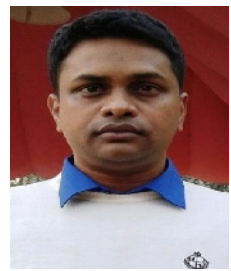

**Md. Yeakub Ali** has completed B.Sc Engineering degree in Electronics & Telecommunication Engineering (ETE) from Rajshahi University Engineering & Technology (RUET) in January 2014. Now, he is working as an Engineer in a telecommunication company. His research area includes Wireless & Mobile Communication, Satellite Communication & Radar.

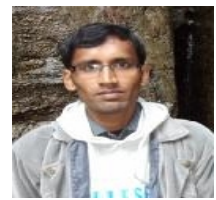

**S. M. Mahmud Hasan** was born in the Rajshahi, the northern city of Bangladesh on 30 January 1991. He has completed B.Sc Engineering degree in Electronics & Telecommunication Engineering (ETE) from Rajshahi University Engineering & Technology (RUET) in September 2012. Now, he is working in the Bangladesh Government Power Development Board (BPDB) as an Assistant Engineer. His research interest includes advanced wireless communication (LTE, LTE-A) etc.

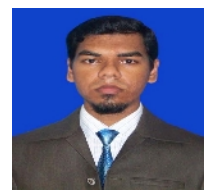